# Surrounding Effects on the Evaporation Efficiency of a Bi-layered Structure in Solar Steam Generation: a Numerical Study

Jinxin Zhong[1], Congliang Huang[1,2a)], Dongxu Wu[1]

[1] School of Electrical and Power Engineering, China University of Mining and Technology, Xuzhou 221116, P. R. China.
[2] Department of Mechanical Engineering, University of Colorado, Boulder, Colorado 80309-0427, USA.
[a)] Author to whom correspondence should be addressed. E-mail: huangcl@cumt.edu.cn.

**Abstract:** The bi-layered structure has drawn a wide interest due to its good performance in solar steam generation. In this work, we firstly develop a calculation model which could give a good prediction of experimental results. Then, this model is applied to numerically study the effects of the depth of the liquid water, the temperature of the ambient air, the temperature of the liquid water, the porosity and the thermal conductivity of the second-layer porous material on the evaporation efficiency. Results show that when the depth of the liquid water is large enough, the thermal insulation at the bottom of the liquid water is not needed. There is a linear dependence of the evaporation efficiency on the temperature of the ambient air or/and the temperature of the liquid water, and an equation has been given to describe this phenomenon in the text. Compared to the temperature of the ambient air, the temperature of the liquid water could have a much larger effect on the evaporation efficiency. The effective thermal conductivity of the second layer, which could impose important effect on the evaporation efficiency, mainly depends on the porosity rather than the thermal conductivity of the second-layer porous material. Thus, we do not need to take into consideration of the thermal conductivity when selecting second-layer materials. This study is expected to provide some information for designing a high-evaporation-performance bi-layered system.

## Nomenclature

| Symbol | Description | Symbol | Description |
|---|---|---|---|
| $a_w$ | Water activity | $R$ | Reaction rate, mol·m$^{-2}$·s$^{-1}$ |
| $c$ | Concentration, mol·m$^{-3}$ | $S$ | Saturation |
| $c_{sat}$ | Saturation vapor concentration, mol·m$^{-3}$ | $S_{iw}$ | Initial water saturation of second layer |
| $c_{v0}$ | Initial water vapor concentration, mol·m$^{-3}$ | $T$ | Temperature, K |
| $C_p$ | Heat capacity, J·kg$^{-1}$·K$^{-1}$ | $\boldsymbol{u}$ | Velocity, m·s$^{-1}$ |
| $D_{cap}$ | Capillary diffusivity, m$^2$·s$^{-1}$ | $\boldsymbol{u}_{mean}$ | Fluid velocity in second layer, m·s$^{-1}$ |
| $D_{eff}$ | Effective vapor diffusivity, m$^2$·s$^{-1}$ | **Greek letters** | |
| $D_{va}$ | Air-vapor diffusivity, m$^2$·s$^{-1}$ | $\varphi_p$ | Porosity |
| $e_b$ | Blackbody hemispherical total emissive power, W·m$^{-2}$ | $\mu$ | Viscosity, kg·m$^{-1}$·s$^{-1}$ |
| $G$ | Irradiation, W·m$^{-2}$ | $\eta$ | Evaporation efficiency, % |
| $H_{vap}$ | Latent heat of evaporation, J·kg$^{-1}$ | $\varepsilon_a$ | Surface absorptivity |
| $\boldsymbol{I}$ | Identity matrix | $\varepsilon_r$ | Surface reflectivity |
| $k$ | Thermal conductivity, W·m$^{-1}$·K$^{-1}$ | $\varepsilon_t$ | Surface transmissivity |
| $k_e$ | Effective Thermal conductivity, W·m$^{-1}$·K$^{-1}$ | $\kappa$ | Permeability of the porous matrix, m$^2$ |
| $K$ | Evaporation coefficient, l·s$^{-1}$ | $\kappa_r$ | Relative permeability, m$^2$ |
| $L_{in}$ | Entrance length, m | $(\rho C_p)_e$ | Effective volumetric heat capacity, J·m$^{-3}$·K$^{-1}$ |
| $M$ | Molecular weight, kg·mol$^{-1}$ | $\rho$ | Density, kg·m$^{-3}$ |
| $\boldsymbol{n}$ | Normal vector | **Subscripts** | |
| $\boldsymbol{n}$ | Mass flux in Eq.(20), kg·m$^{-2}$·s$^{-1}$ | 0 | Ambient air |
| $P$ | Pressure, Pa | a | Dry air |
| $P_{in}$ | Entrance pressure, Pa | bottom | Lower surface of bi-layer structure |
| $P_{out}$ | Exit pressure, Pa | g | Gas in second layer |
| $P_{set}$ | Saturation pressure in second layer, Pa | p | Porous second layer |
| $\boldsymbol{q}$ | Heat flux vector, W·m$^{-2}$ | fside | Front side of bi-layer structure |
| $\boldsymbol{q}_e$ | Radiation heat flux, W·m$^{-2}$ | rside | Rear side of bi-layer structure |
| $q_{eff}$ | Efficient thermal, W·m$^{-2}$ | tot | Fluid of second layer |
| $q_l$ | Thermal loss, W·m$^{-2}$ | top | Top surface of bi-layer structure |
| $q_{tot}$ | Total thermal input, W·m$^{-2}$ | v | Air in ambient air domain |
| $Q_{vap}$ | The heat of evaporation, W·m$^{-3}$ | w | Water in second layer |
| $R$ | Ideal gas constant in Eq.(6), J·mol$^{-1}$·K$^{-1}$ | wg | Vapor in second layer |

## 1. Introduction

Contrasting to the traditional evaporation process which requires the consumption of fossil energy [1-4], solar-enabled evaporation is green [5-7] which has a potential application in wastewater treatment [8], desalination [9], medical sterilization [10] and even for electricity generation [11]. Many efforts have been dedicated to minimize the heat losses and to improve the efficiency of water evaporation by using a bi-layered structure for solar steam generation which was firstly reported by Ghasemi et al. [12]. The absorbent material is usually applied as the first layer of a bi-layered structure to absorb sunlight, when the porous material is commonly used as the second layer for water absorption and heat insulation [13-15]. Numbers of efforts have been dedicated to find novel first-layer materials for high efficiency of light absorbing, such as nanoparticle [16-18], activated carbon[19], carbon nanotubes [20], polypyrrole [9], and graphene [21]. The second-layer material also has drawn a wide interest, and the organic porous material is usually applied as the second-layer material because of high capillary-effect performance and low thermal conductivity, such as wood [22, 23], polyvinyl alcohol [9], carbonized mushrooms [24], and cellulose nanofibers [25]. Although a large number of studies have been carried out to probe cheap and efficient first or/and second layer materials to increase the evaporation efficiency of the bi-layered structure, to our knowledge, there is scarcely systematical study to probe proper micro-structures of materials to enhance the evaporation rate. The influence of the ambient environment on the evaporation efficiency is also scarcely considered in previous works, when the depth and the temperature of the liquid water could impose important effects on the evaporation efficiency.

    In this work, we firstly develop a calculation model for simulating the evaporation efficiency of the bi-layered structure, then this model is verified by the experiments. Finally, the model is used to probe the effects of the temperature of the ambient air, the depth and temperature of the liquid water on the evaporation efficiency. Results show that the calculation model could give a good prediction of experiments, and when the depth of the liquid water is large enough, the thermal insulation at the bottom of the liquid water is not needed. The evaporation efficiency increases linearly with the increase of the temperature of the ambient air or/and the temperature of the liquid water. This study is expected to provide some information for designing a high-evaporation-performance bi-layered system.

## 2. Mathematical model

### 2.1 Physical model

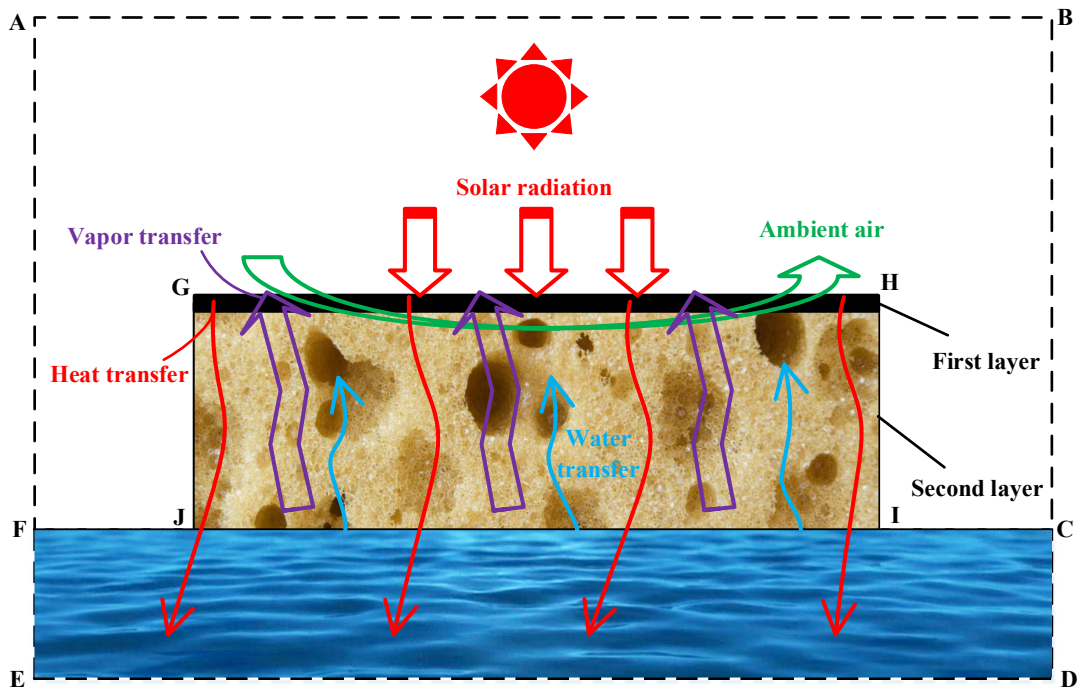

*Fig. 1 Schematic bi-layered system for solar steam generation, where the system is composed of the bi-layered structure and its surroundings including the liquid water and the ambient air.*

The water evaporation in a bi-layered system is schematically shown in Fig. 1. In such a system, the solar radiation causes a high temperature at the first layer, and this high temperature further heats the second layer which is composed of a porous material and the water holding in it. The high temperature of the second layer could generate vapor, and the capillary effect of the second layer makes sure that the liquid water could flow into the second layer to replenish the water losses caused by the evaporation. In this evaporation process, the ambient air could flow over the upside of the first layer, and the air could diffuse into the second layer and takes away the vapor. It is clear that there could be three different physical processes, including heat transfer, vapor transfer and water transfer. To analyze such a system, we make the following assumptions:

**For the first layer:**
(1) Considering the thickness of the first layer has negligible effect on the evaporation rate, here we ignore the thickness of the first layer and regard it as a diffuse surface;
(2) Ignoring the thermal loss through the thermal radiation of the first layer to the environment;

**For the second layer:**
(3) Ignoring the thermal loss through the thermal radiation of the second layer to the environment;
(4) The structure of the second layer is isotropic;
(5) The viscosity dissipation like heat transfer and work caused by pressure changing is not considered to meet the assumption of local thermal equilibrium;
(6) The vapor is in equilibrium with the liquid phase, in other words, the time scale of

the evaporation is much smaller than the time scale of the transport phenomena like heat and mass transfer;

(7) The second layer absorbs water only from its bottom surface;

**For the ambient air:**

(8) The ambient-air flow is assumed to be laminar because of their small Reynolds number.

## 2.2 Governing equations

Based on above assumptions, the governing equations for each part of the system are summarized in this section, and the coupling of different governing equations are illustrated in Fig. 2.

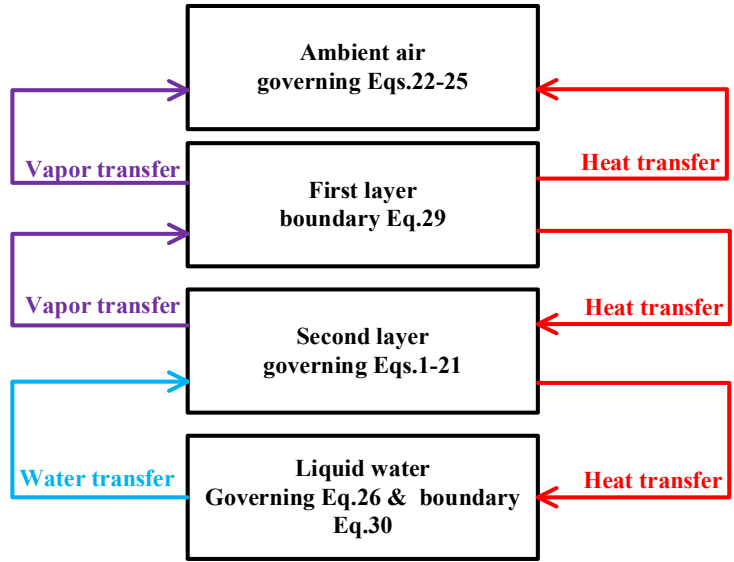

*Fig. 2 The heat and mass transfer in a bi-layered system.*

### 2.2.1 The second layer

The air flow in the second layer will determine the pressure gradient of gas phase, and the pressure gradient accompanying with capillary effect are crucial for the water transfer. The water, air and vapor could fill all of the pores in the second layer, thus the saturation concentrations of the gas (air and vapor) and liquid phases (water) must satisfy the saturation constraint equation. In the following part, the coupled liquid and gas transfer are respectively introduced, and finally the heat transfer equation is given.

**Saturation constraint equation:**

$$S_g + S_w = 1. \tag{1}$$

where $S_g$ is the gas saturation, and $S_w$ is the saturation of water can be calculated by,

$$S_w = \frac{c_w M_w}{\rho_w \varphi_p}. \tag{2}$$

where $c_w$ is the concentration of water in the second layer, which can be derived from the water transport equation.

**Water transport equation:**

$$\frac{\partial c_w}{\partial t} + \nabla \cdot (-D_{cap} \nabla c_w) + \boldsymbol{u}_w \cdot \nabla c_w = R_w, \tag{3}$$

where the capillary diffusivity $D_{cap}$ can be empirically or experimentally obtained [26], and the water velocity $\boldsymbol{u}_w$ is calculated by Darcy's Law which is defined in terms of the gas-phase pressure gradient $\nabla P_g$,

$$\boldsymbol{u}_w = -\frac{\kappa \kappa_{rw}}{S_w \varphi_p \mu_w} \nabla P_g \boldsymbol{I}, \tag{4}$$

where the overall permeability $\kappa$ depends on the porosity of the material which will be carefully considered in part 3 of this paper. The liquid relative permeability $\kappa_{rw}$ can be empirically or experimentally obtained [26].

Because mass is balanced between the liquid water and steam in the second layer, the reaction rate of liquid water $R_w$ in Eq. (3) is the opposite of the rate of steam generation in the second layer,

$$R_w = -K \cdot (a_w c_{sat} - c_{wg}), \tag{5}$$

where the water activity $a_w$ describes the amount of water that evaporates into air depending on both of the water content and the temperature in the surrounding air [26]. The saturation concentration $c_{sat}$ is determined by both of the saturation pressure and the temperature of the second layer,

$$c_{sat} = \frac{P_{set}}{RT_p}, \tag{6}$$

where the saturated vapor pressure $P_{set}$ can be described by Tentens formula,

$$P_{set} = 610.7 \times 10^{\frac{7.5 \times (T_p - 273.15)}{T - 35.85}}. \tag{7}$$

**Vapor transport equation:**

$$\frac{\partial c_{wg}}{\partial t} + \nabla \cdot (-D_{eff} \nabla c_{wg}) + \boldsymbol{u}_{wg} \cdot \nabla c_{wg} = R_{wg}, \tag{8}$$

where the effective diffusion coefficient of vapor $D_{eff}$ can be described by the Millington and Quirk equation,

$$D_{eff} = D_{va} S_g^{\frac{10}{3}} \varphi_p^{\frac{4}{3}}. \tag{9}$$

In Eq. (8), the velocity field of vapor $\boldsymbol{u}_{wg}$ is superposed by the velocity fields derived from the convection diffusion and vapor diffusion,

$$\boldsymbol{u}_{wg} = \frac{u_g}{S_g \varphi_p} - \frac{M_v D_{eff}}{M_g \rho_g} \nabla \rho_g. \tag{10}$$

In Eq. (8), the rate of steam generation $R_{wg}$ is determined by the saturation concentration $c_{sat}$ and water activity $a_w$,

$$R_{wg} = K \cdot (a_w c_{sat} - c_{wg}). \tag{11}$$

**Gas (vapor + air) continuity equation:**

$$\nabla \cdot (\rho_g \boldsymbol{u}_g) = 0. \tag{12}$$

**Gas momentum conservation equations:**

$$\frac{\rho_g}{\varphi_p^2} \boldsymbol{u}_g \cdot \nabla \boldsymbol{u}_g = -\nabla P_g \boldsymbol{I} + \nabla \cdot \left[ \frac{1}{\varphi_p} \left\{ \mu_g (\nabla \boldsymbol{u}_g + (\nabla \boldsymbol{u}_g)^T) - \frac{2}{3} \mu_g (\nabla \cdot \boldsymbol{u}_g) \boldsymbol{I} \right\} \right] - (\kappa \kappa_{rg})^{-1} \mu_g \boldsymbol{u}_g. \quad (13)$$

In Eqs. (12) and (13), the thermo-physical properties of gas is obtained by mixing the property of vapor and dry air.

**Energy conservation equation:**

$$(\rho C_p)_e \frac{\partial T}{\partial t} + \rho_{tot} C_{p,tot} \boldsymbol{u}_{mean} \cdot \nabla T - k_e \nabla^2 T = Q_{vap}, \quad (14)$$

where the thermo-physical properties of fluids (gas and water) are calculated by,

$$\rho_{tot} = S_g \rho_g + S_w \rho_w, \quad (15)$$

$$C_{p,tot} = \frac{S_g \rho_g C_{p,g} + S_w \rho_w C_{p,w}}{\rho_{tot}}, \quad (16)$$

$$k_{tot} = S_g k_g + S_w k_w. \quad (17)$$

In Eq.(14), the effective thermal properties of the porous material are described using the following equations,

$$(\rho C_p)_e = \varphi_p \rho_{tot} C_{p,tot} + (1 - \varphi_p) \rho_p C_{p,p}, \quad (18)$$

$$k_e = \varphi_p k_{tot} + (1 - \varphi_p) k_p. \quad (19)$$

The velocity is averaged from that of moist air, water vapor and liquid water,

$$\boldsymbol{u}_{mean} = \frac{\boldsymbol{n}_a C_{p,a} + \boldsymbol{n}_{wg} C_{p,wg} + \boldsymbol{n}_w C_{p,w}}{\rho_{tot} C_{p,tot}}, \quad (20)$$

where $\boldsymbol{n}_a$, $\boldsymbol{n}_{wg}$ and $\boldsymbol{n}_w$ is respectively the mass flux of dry air, vapor and water [27]. Because the evaporation is an endothermic process, the heat in vaporization in Eq. (14) is described by,

$$Q_{vap} = -H_{vap} M_w R_{wg}. \quad (21)$$

### 2.2.2 Ambient air

Continuity equation:

$$\nabla \cdot (\rho_v \boldsymbol{u}_v) = 0. \quad (22)$$

Momentum conservation equations:

$$\rho_v (\boldsymbol{u}_v \cdot \nabla) \boldsymbol{u}_v = \nabla \cdot \left[ -P_v \boldsymbol{I} + \mu_v (\nabla \boldsymbol{u}_v + (\nabla \boldsymbol{u}_v)^T) - \frac{2}{3} \mu_v (\nabla \cdot \boldsymbol{u}_v) \boldsymbol{I} \right]. \quad (23)$$

Energy conservation equation:

$$\rho_v C_{p,v} \frac{\partial T}{\partial t} + \rho_v C_{p,v} \boldsymbol{u}_v \cdot \nabla T - k_v \nabla^2 T = 0. \quad (24)$$

Vapor transport equation:

$$\frac{\partial c_v}{\partial t} + \nabla \cdot (-D_{va} \nabla c_v) + \boldsymbol{u}_v \cdot \nabla c_v = 0. \tag{25}$$

In Eqs. (22)-(25), the thermo-physical properties of gas is obtained by mass-weighted averaging the property of vapor and dry air.

### 2.2.3 Liquid water

Energy conservation equation:

$$\rho_w C_{p,w} \frac{\partial T}{\partial t} - k_w \nabla^2 T = 0. \tag{26}$$

### 2.3 Boundary conditions

The boundary conditions for the air flow, heat transfer, water transfer, and vapor transport are respectively given in the following parts, and also summarized in Table 1.

**For air flow:** FA in Fig. 1(a) is the inlet boundary defined by the Eq. (27), and BC is the outlet boundary defined by the Eq. (28),

$$L_{in} \nabla \cdot \left[ -P_v \boldsymbol{I} + \mu_v (\nabla \boldsymbol{u}_v + (\nabla \boldsymbol{u}_v)^T) - \frac{2}{3} \mu_v (\nabla \cdot \boldsymbol{u}_v) \boldsymbol{I} \right] = -P_{in} \boldsymbol{n}, \tag{27}$$

$$\left[ -P_v \boldsymbol{I} + \mu_v (\nabla \boldsymbol{u}_v + (\nabla \boldsymbol{u}_v)^T) - \frac{2}{3} \mu_v (\nabla \cdot \boldsymbol{u}_v) \boldsymbol{I} \right] = -P_{out} \boldsymbol{n}. \tag{28}$$

**For heat transfer:** Considering that BC is the air outflow boundary, only will the heat transfer via air convection happens there. FA is the fixed temperature boundary where the temperature is fixed as that of the inflow air. GH is set as a diffuse surface, and a radiative heat source which is defined by the Eq. (29) is added on it,

$$-\boldsymbol{n} \cdot \boldsymbol{q}_e = \varepsilon_a (G - e_b). \tag{29}$$

**For water transport:** The pores at the IJ boundary are full filled with water, and therefore the condition at the IJ boundary could be defined by,

$$c_w = S_{w0} \varphi_p \rho_w M_w^{-1}. \tag{30}$$

**For vapor transfer:** The inlet vapor concentration at FA is defined as $c_{v0}$. BC is the vapor outlet boundary.

**Table 1** Boundary conditions.

| Boundaries | Air flow | Heat transfer | Water transport | Vapor transport |
|---|---|---|---|---|
| AB | $\boldsymbol{u} = 0$ | $-\boldsymbol{n} \cdot \boldsymbol{q} = 0$ | / | $-\boldsymbol{n}(-D_{va} \nabla c_v + \boldsymbol{u}_v c_v) = 0$ |
| BC | Eq.(28) | $-\boldsymbol{n} \cdot \boldsymbol{q} = 0$ | / | $-\boldsymbol{n} \cdot D_{va} \nabla c_v = 0$ |
| CD | / | $-\boldsymbol{n} \cdot \boldsymbol{q} = 0$ | / | / |
| DE | / | $-\boldsymbol{n} \cdot \boldsymbol{q} = 0$ | / | / |
| EF | / | $-\boldsymbol{n} \cdot \boldsymbol{q} = 0$ | / | / |
| FA | Eq.(27) | $T = T_0$ | / | $c_v = c_{v0}$ |
| GH | Coupling boundary | Eq.(29) | $-\bm{n}(-D_{cap}\nabla c_w + \bm{u}_w c_w) = 0$ | Coupling boundary |
| HI | Coupling boundary | Coupling boundary | $-\bm{n}(-D_{cap}\nabla c_w + \bm{u}_w c_w) = 0$ | Coupling boundary |
| IJ | $\bm{u} = 0$ | Coupling boundary | Eq.(30) | $-\bm{n}(-D_{eff}\nabla c_{wg} + \bm{u}_{wg} c_{wg}) = 0$ |
| JG | Coupling boundary | Coupling boundary | $-\bm{n}(-D_{cap}\nabla c_w + \bm{u}_w c_w) = 0$ | Coupling boundary |
| JF | $\bm{u} = 0$ | Coupling boundary | / | $-\bm{n}(-D_{va}\nabla c_v + \bm{u}_v c_v) = 0$ |
| CI | $\bm{u} = 0$ | Coupling boundary | / | $-\bm{n}(-D_{va}\nabla c_v + \bm{u}_v c_v) = 0$ |

## 3. Numerical simulation

The COMSOL Multiphysics software is used to numerically solve the governing equations, where the User-Defined Variable (UDV) of material properties and the User-Defined Function (UDF) of source terms are self-developed according to the calculation models in part 2. The following sizes is applied in the simulation: the lengths of AB, BC, CD, GH and HI is respectively 120, 50, 40, 30 and 10 mm. The overall permeability of the second-layer porous material is obtained by numerically solving the Creeping flow model by finite element method performed in a geometrical model. The geometrical model is constituted by arranging two different-size spheres in a regular body-centered cubic (bcc) structure. By changing the radius of the spheres in the bcc structure, the permeability at different porosity is obtained. Substituting the porosity in the creeping flow model, the mass flow can be calculated, then the result is compared with that predicted by Darcy model (Darcy's law) where the porosity and its resultant permeability should be substituted in. Fig. 3(a) shows that there is a good agreement between creeping flow model and the Darcy model suggesting that the obtained permeability could capture the structure well. The permeability at different porosities is shown in Fig. 3 (b). These permeability will be applied in Eqs. (4) and (13) in the numerical simulations. Other material properties and parameters used in the simulation are summarized in Table 2.

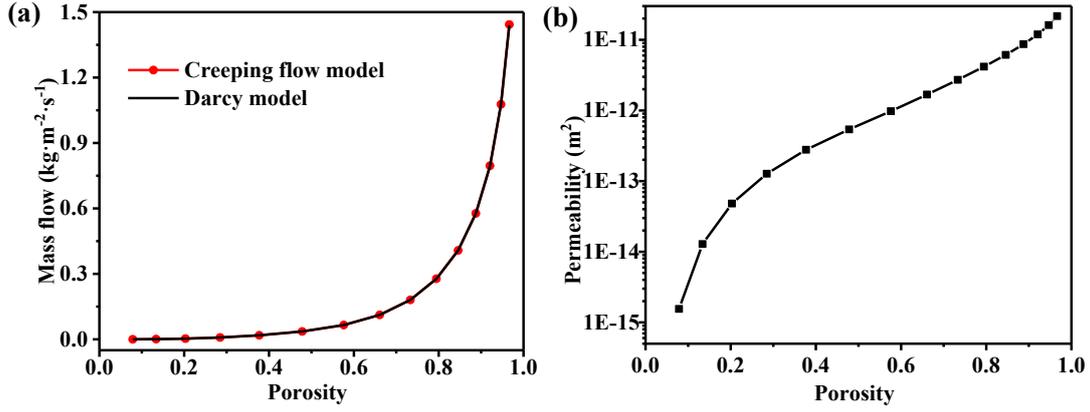

Fig. 3 Effect of the porosity on the permeability: (a) comparison between the Creeping flow model and the Darcy model; (b) permeability of the second layer at different porosities.

Table 2 Material properties and parameters used in the simulation.

| Nomenclature | Value |
| --- | --- |
| Initial ambient pressure, $P_0$ | 1 atm. |
| Initial ambient air temperature, $T_0$ | 293.15 K |
| Initial liquid water temperature, $T_w$ | 293.15 K |
| Initial ambient air velocity, $\boldsymbol{u}_0$ | 0.4 m·s$^{-1}$ |
| Initial ambient vapor concentration, $c_{v0}$ | 0.3357 mol·m$^{-3}$ |
| Radiation heat flux, $q_e$ | 1200 W·m$^2$ |
| Molecular weight of dry air, $M_a$ | 0.028 kg·mol$^{-1}$ |
| Dry air viscosity, $\mu_a$ | 1.81×10$^{-5}$ kg·m$^{-1}$·s$^{-1}$ |
| Dry air thermal conductivity, $k_a$ | 0.025 W·m$^{-1}$·K$^{-1}$ |
| Dry air heat capacity, $C_{p,a}$ | 1.006×10$^{-5}$ J·kg$^{-1}$·K$^{-1}$ |
| Dry air density, $\rho_a$ | 1.205 kg·m$^{-3}$ |
| Molecular weight of water, $M_w$ | 0.018 kg·mol$^{-1}$ |
| Water viscosity, $\mu_w$ | 1.002×10$^{-3}$ kg·m$^{-1}$·s$^{-1}$ |
| Water thermal conductivity, $k_w$ | 0.59 W·m$^{-1}$·K$^{-1}$ |
| Water heat capacity, $C_{p,w}$ | 4.182×10$^3$ J·kg$^{-1}$·K$^{-1}$ |
| Water density, $\rho_w$ | 998.2 kg·m$^{-3}$ |
| Vapor viscosity, $\mu_{wg}$ | 1.8×10$^{-5}$ kg·m$^{-1}$·s$^{-1}$ |
| Vapor thermal conductivity, $k_{wg}$ | 0.026 W·m$^{-1}$·K$^{-1}$ |
| Vapor heat capacity, $C_{p,wg}$ | 2.062×10$^3$ J·kg$^{-1}$·K$^{-1}$ |
| Porosity of second layer, $\varphi_p$ | 0.9 |
| Thermal conductivity of porous material, $k_p$ | 0.14 W·m$^{-1}$·K$^{-1}$ |
| Heat capacity of porous material, $C_{p,p}$ | 1650 J·kg$^{-1}$·K$^{-1}$ |
| Density of porous material, $\rho_p$ | 800 kg·m$^{-3}$ |
| Evaporation coefficient, $K$ | 100000 1·s$^{-1}$ |
| Ideal gas constant, $R$ | 8.314 J·mol$^{-1}$·K$^{-1}$ |
| Surface absorptivity, $\varepsilon_a$ | 0.9 |
| Air-vapor Diffusivity, $D_{va}$ | 2.6×10$^{-5}$ m$^2$·s$^{-1}$ |
| Latent heat of evaporation, $H_{vap}$ | $2.454 \times 10^6$ J·kg$^{-1}$ |
| Initial water saturation of second layer, $S_{iw}$ | 1 |

Appling customized mesh in boundary layers, the evaporation rate for different number of mesh grids are studied. The increase of the number of grids will have a negligible effect on the results when the number of grids becomes larger than 36650. In this work, the mesh with 36650 grids is applied. Fig. 5 demonstrates that the calculation models in this work could give reasonable velocity, temperature, vapor concentration and relative humidity fields. To analyze the thermal losses, the heat conserved from the absorbed solar energy by the first layer is expressed as,

$$q_{tot} = q_{eff} + q_{l,top} + q_{l,bottom} + q_{l,fside} + q_{l,rside}, \qquad (31)$$

where $q_{eff}$ stand for the effective heat used to evaporate water, $q_l$ represents the thermal losses, the subscript 'top', 'bottom', 'fside' and 'rside' respectively signifies the top surface, bottom surface, windward surface (front side) and leeward surface (rear side) of the bi-layered structure. The thermal losses of each part are given in the Fig. 5(c). The superscript '*' in this paper indicates a dimensionless quantity scaled by that of the total heat $q_{tot}$.

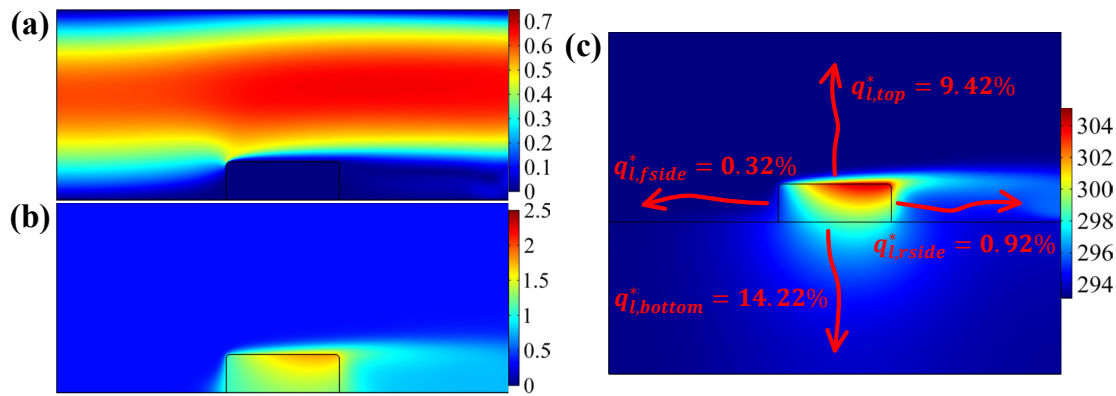

*Fig. 4 Simulation results: (a) velocity field, with the unit of m·s$^{-1}$; (b) vapor concentration field, with the unit of mol·m$^{-3}$; and (c) temperature field with the unit of K and the scaled thermal losses by the total heat.*

## 4. Results and discussions

The calculation model is firstly verified by experiments, and then the model is used to probe the effects of some influence factors on the evaporation efficiency, including the depth of the liquid water, the temperature of the surroundings, the porosity and the thermal conductivity of the second-layer porous materials. To understand the evaporation efficiency from the angle of energy conservation, the thermal losses dissipating through different surface of the bi-layered structure are also respectively analyzed.

### 4.1 Experiment validation

**1) Material preparation and characterization**

The bi-layered structure is prepared by daubing 50-nm-diameter copper particles on a cellulose sponge, where the copper nanoparticles act as the first layer and the sponge is used as the second layer. The daubing density of copper nanoparticles is applied as 37.3 g·m$^{-2}$ for an efficient light absorbing (approximately 99%). Copper particles are commercially obtained from Beijing DK Nano technology Co., Ltd, and the cellulose sponge is made from wood fiber. Microstructures of the first layer and the second layer is shown in Fig. 6, observed with field-emission scanning electron microscopy (SEM) on a QuantaTM-250 instrument (FEI Co., USA) with an accelerating voltage of 30 or 25 kV. It shows that the copper particles distribute evenly on the surface of the sponge which could make sure an efficient light absorbing.

The absorptivity of the first layer can be calculated by,

$$\varepsilon_a = 1 - \varepsilon_r - \varepsilon_t, \tag{32}$$

where $\varepsilon_t$ and $\varepsilon_r$ are respectively the surface transmissivity and reflectivity. The $\varepsilon_r$ and the $\varepsilon_t$ are measured in a range of 200-1200 nm with a Spectrophotometer (Lambda 750S, PerkinElmer Inc.). The calculated absorptivity is shown in Fig. 7. The thermal conductivity of the second-layer porous material with the filling air extruded out is measured to be about 0.1452 W·m$^{-1}$·K$^{-1}$ with the transient hot-wire method performed on a commercial device (TC3200, Xian XIATECH Technology Co.)[28, 29]. The porosity of the second layer is calculated by mass-volume method, where the volume is measured with Archimedes drainage method. The porosity is calculated to be about 0.92.

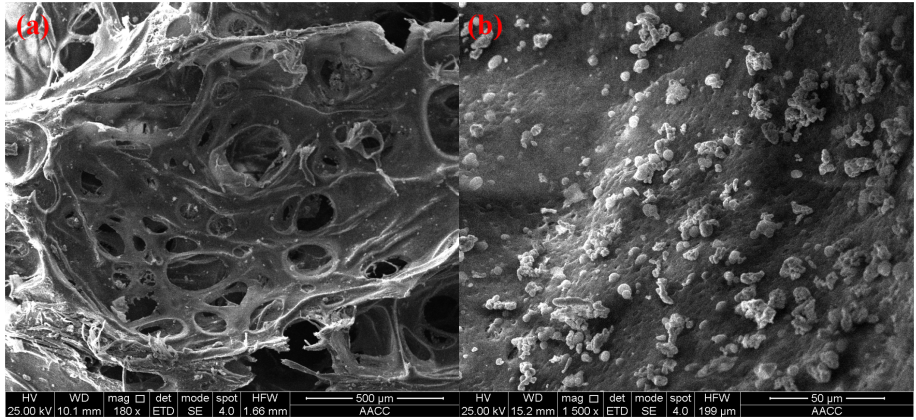

*Fig. 5 Microstructures of materials used in the bi-layered system: (a) pores in cellulose sponge; and (b) Top surface of cellulose sponge daubed with 50-nm copper particles.*

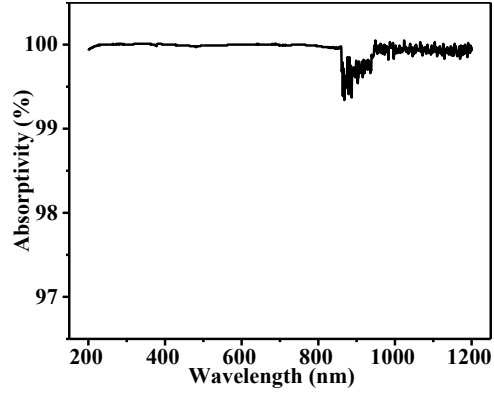

*Fig. 6 Absorptivity of the first-layer calculated by Eq. 32.*

**2) Experiment and Results**

A solar simulator (BOS-X-350G, Bosheng Quantum Technology Co., Ltd.) is used to supply solar radiation, and a data acquisition unit (Model 2700, Keithley Instruments, Inc.) is used to collect the temperature. The mass change of the water during the evaporation process is obtained by an electronic analytical balance (CP214, OHAUS Instruments (Shanghai) Co., Ltd.). The conditions applied in the experiment are summarized in Table 3. Five separate experiments (No. 1 to 5) are performed on the same bi-layered structures at the same conditions. The experimental temperature rise and evaporation rate are respectively shown in Fig. 9(a) and (b), and our numerical results are also added in figures for comparison. The deviation of the numerical results is less than 5 % compared to that of experiments, which indicates that the calculation models developed in this work is reasonably soundable.

**Table 3** Conditions in the experiment.

| Parameters | Value |
| --- | --- |
| Radiation heat flux, $q_e$ | 1200 W·m² |
| Surface absorptivity, $\varepsilon_a$ | 0.99 |
| Thermal conductivity of porous material, $k_p$ | 0.1452 W·m$^{-1}$·K$^{-1}$ |
| Porosity of porous material, $\varphi_p$ | 0.92 |

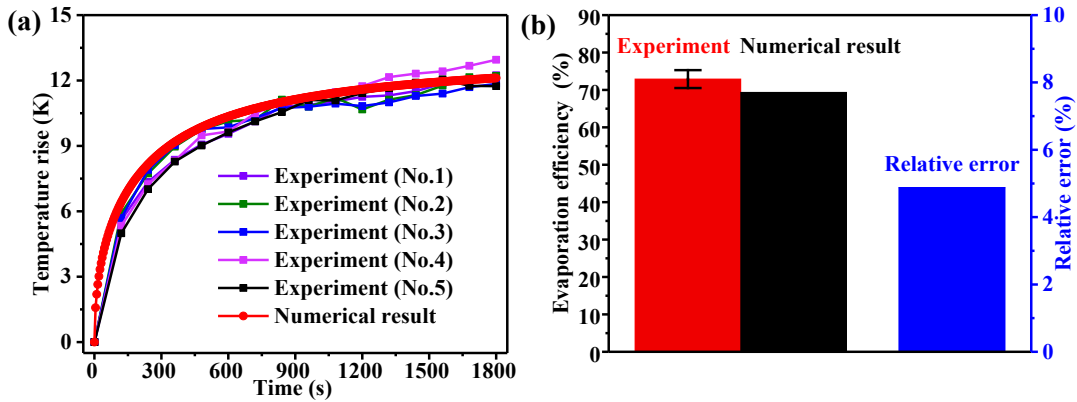

*Fig. 7 Results comparison between experimental and numerical methods: (a) the temperature rise at the top surface of the second layer; (b) the relative error and the*

*evaporation efficiency.*

## 4.2 Influence of the liquid water depth

With the depth of the liquid water increased from 1 to 41 mm, the evaporation efficiency initially decreases sharply and then gradually converges to a constant value of about 64.5 % after the liquid-water depth reaching 20mm, as depicted in Fig. 10(a). Because more water can depress the thermal-loss-caused temperature rise of the liquid water as illustrated in the inset in Fig. 10(a), the lower temperature of the liquid water could enlarge the temperature difference between the second layer and the water. This enlarged temperature difference leads to the escalation of the thermal loss from the bottom surface of the second layer ($q_{l,bottom}$), as illustrated in Fig. 10(b). The increased thermal loss further lead to the sharp decrease of the evaporation efficiency in Fig. 10(a). When the liquid water is more enough, the thermal loss from the bottom surface of the second layer will be limited by the convective heat-transfer coefficient of the water. Thus, the evaporation efficiency finally saturates to a constant value. It can be concluded that when the depth of the liquid water is large enough, the thermal insulation at the bottom of the liquid water is not needed. In this work, this depth is about 20 mm.

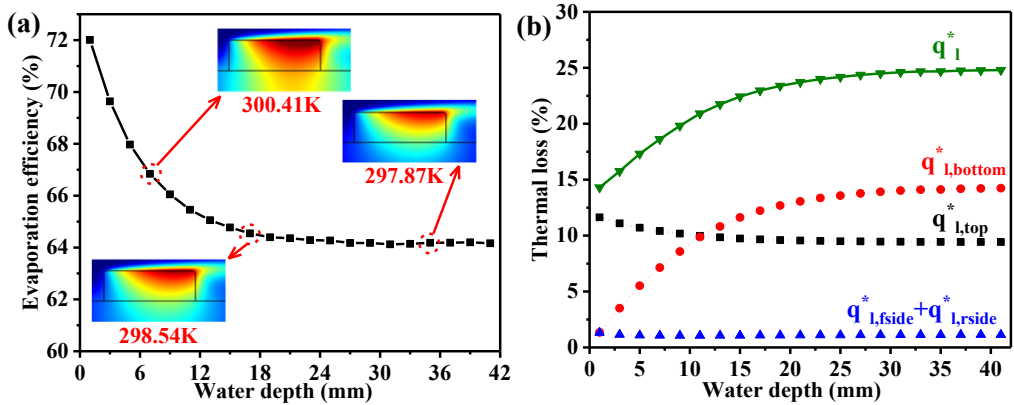

*Fig. 8 Influence of the liquid water depth: (a) evaporation efficiency; (b) thermal losses.*

## 4.3 Influence of temperatures of the ambient air and the liquid water

The heat conserved from the solar energy absorbed by the first layer ($Q_{tot}$) can be divided into three parts: the effective heat used to evaporate water ($Q_{eff}$), the thermal losses dissipating to the ambient air ($Q_{l,top} + Q_{l,fside} + Q_{l,rside}$) and the thermal losses dissipating to the liquid water through the bottom surface of the second layer ($Q_{l,bottom}$). Because the thermal losses dissipating to the liquid water and to the ambient air is mainly due to the convective heat transfer, the total heat can be expressed as,

$$Q_{tot} = Q_{eff} + (Q_{l,top} + Q_{l,fside} + Q_{l,rside}) + Q_{l,bottom} \\ \approx A_{top}q_{eff} + A_{top}h_0(T_1 - T_0) + A_{bottom}h_w(T_2 - T_w), \quad (33)$$

where $A$ stand for the area. Thus, the water evaporation can be approximately written as,

$$\eta = \frac{q_{eff}}{q_{tot}} \approx a + bT_0 + cT_w, \quad (34)$$

where $a = \frac{q_{tot} - h_0 T_1 - h_w T_2}{q_{tot}}$, $b = \frac{h_0}{q_{tot}}$, $c = \frac{h_w}{q_{tot}}$. $h_0$ and $h_w$ are respectively the convection heat transfer coefficient of the ambient air and the liquid water. $T_1$ and $T_2$ are respectively the temperatures at interfaces between the system and the ambient air or the liquid water. $T_0$ and $T_w$ are respectively the temperatures of the ambient air and the liquid water. Eq. (34) tells that the evaporation efficiency linearly depends on the $T_0$ and $T_w$. The value of $a$, $b$ and $c$ is obtained by fitting Eq.(34) to the numerical results in this work.

The numerical results of the evaporation efficiency are shown in Figs. 11(a) and (b). Fitting by the least square method, the semi-empirical model about the evaporation efficiency is obtained,

$$\eta \approx -615 + 0.4796 T_0 + 1.839 T_w, \tag{35}$$

The coefficient of determination for the fitting is larger than 0.99, and the root-mean-square error is less than 0.72. The model and the numerical results are both shown in Fig. 11(c) for comparison, and the relative error of the model is further depicted in Fig. 11(d). It shows that the model could perfectly capture the numerical results and the deviation of the model result is less than 2%.

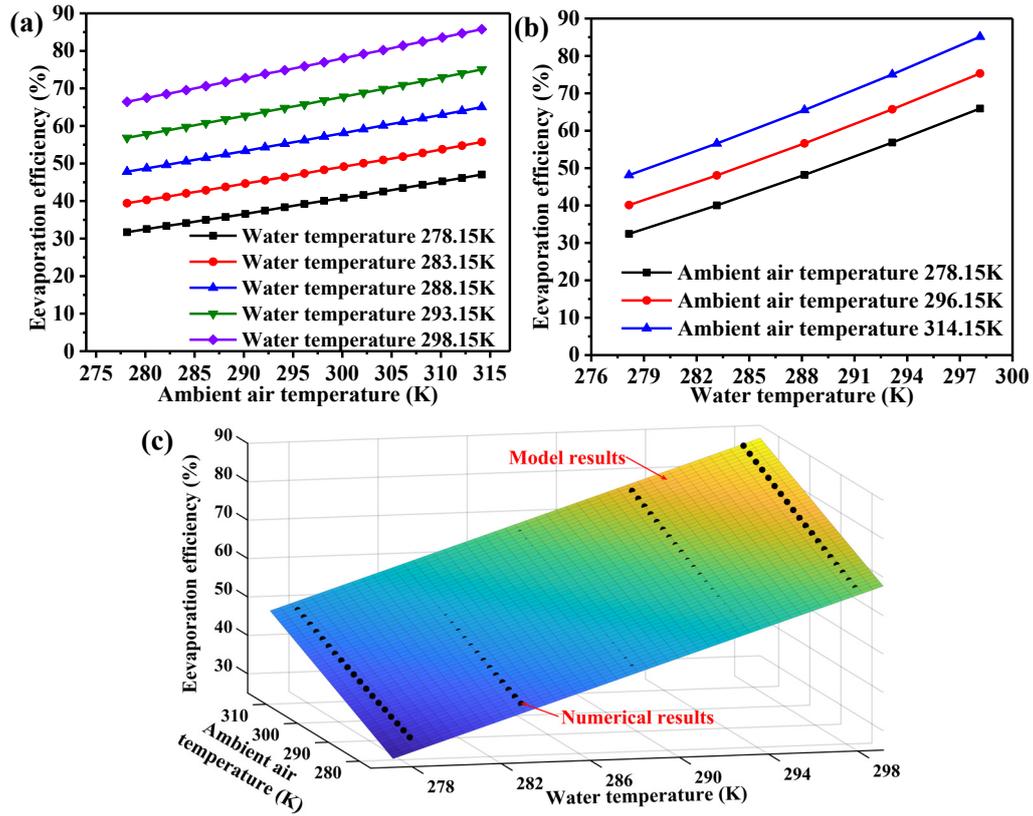

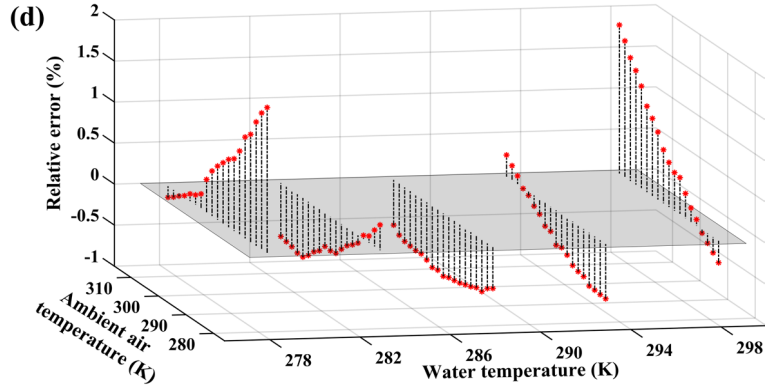

*Fig. 9 Influence of the temperature of the liquid water and the ambient air: (a),(b) evaporation efficiency; (c) comparison between the model and the numerical results; (d) the relative error of the model.*

To reveal the underlying mechanism of the linear increase of the evaporation efficiency, the thermal losses through different surfaces of the bi-layered structure are exhibited in Fig. 12. The thermal losses decrease linearly with the increase of the temperatures of both the ambient air and the liquid water, due to the decreased temperature difference between the bi-layered structure and the ambient air or the liquid water. Eq. (35) also tells that the temperatures of the liquid water could have a much larger effect on the evaporation efficiency with respect to the ambient air. This can be easily understood by that the conversion heat transfer coefficient of the water is much larger than that of the air.

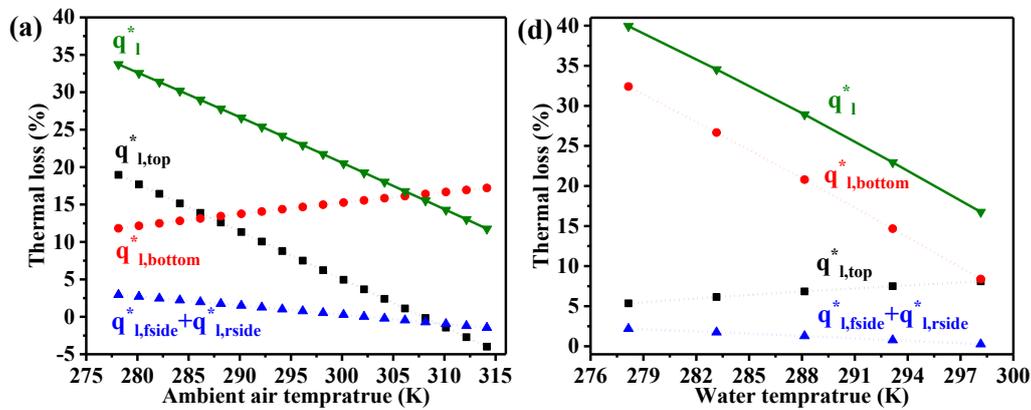

*Fig. 10 The thermal losses: (a) changing ambient-air temperature while maintaining the liquid-water temperature at 293.15K; (b) changing the liquid-water temperature while maintaining the ambient-air temperature at 296.15K.*

### 4.4 Influence of the porosity and the thermal conductivity of the second-layer

Taking into account the porosity-determined permeability and the thermal losses dissipating into the liquid water, the evaporation efficiency is shown in Fig. 13(a). The evaporation efficiency firstly increases rapidly with the porosity increasing from 0.2 to 0.5, and then gradually decreases. There is an optimal porosity for obtaining a maximum evaporation efficiency. To understand the underlying mechanism of the existence of the optimum porosity, the thermal losses from each surface of the bi-layered structure are shown in Fig. 13(b). When a low porosity is applied, there is little water could be hold in pores and thus the effective thermal conductivity of the second layer will be low. This low effective thermal conductivity of the second-layer could result in a higher temperature at the top surface of the bi-layered structure, as illustrated in the inset of the Fig. 13(a). The higher temperature leads to a larger temperature difference between the bi-layered structure and the ambient air or liquid water, the thermal loss therefore is high (Fig. 13b). When a high porosity of about 0.9 is applied, the high effective thermal conductivity of the second layer could lead to a lower temperature of the bi-layered structure, thus the thermal loss is low. The largest evaporation efficiency occurring at the porosity of about 0.55. When the porosity is increased in the second layer, the temperature difference (positive effect) between the bi-layered structure and the ambient will be decreased while the negative effect of the increasing effective thermal conductivity. The balanced influence of the positive and the negative effects should be responsible for the optimum porosity.

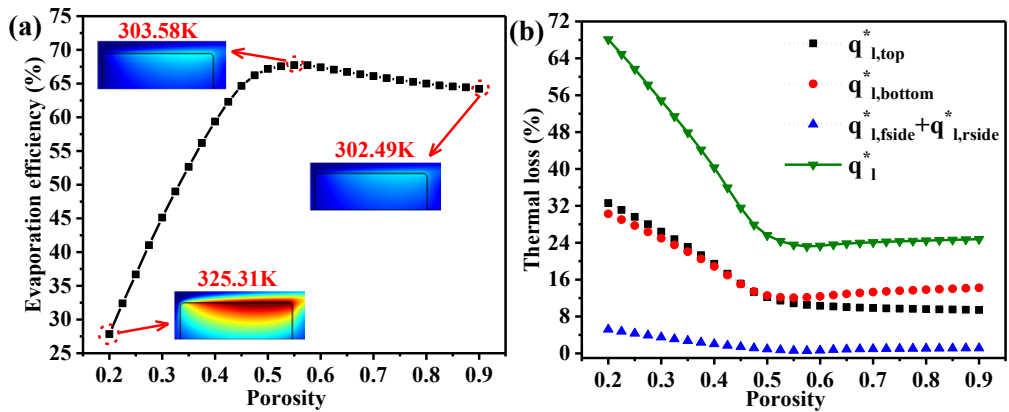

*Fig. 11 Influence of the second-layer porosity: (a) evaporation efficiency; (b) thermal losses with porosity.*

The effect of the thermal conductivity of the porous material on evaporation efficiency is small for all porosity, as depicted in Fig. 14 (a). Analyzing the thermal loss at a porosity of 0.9, the thermal conductivity ($k_p$) of the high porous material in the second layer has a negligible effect on the thermal loss. This quite small effect of the thermal conductivity of the porous material can be easily understood by that: the effective thermal conductivity of the second layer mainly depends on the porosity rather than the thermal conductivity of the porous material when a larger porosity could lead to more water in the second layer. We do not need to take into consideration of the thermal conductivity when selecting the proper second-layer materials.

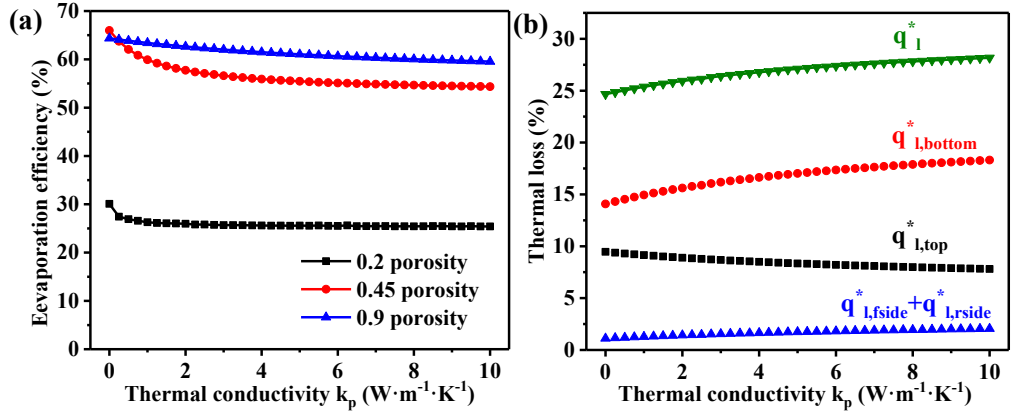

Fig. 12 Influence of the thermal conductivity of the porous material: (a) evaporation efficiency; (b) thermal losses with thermal conductivity of porous material at the porosity of 0.9.

## 5. Conclusions

A calculation model has been developed to analyze the evaporation efficiency of the bi-layered structure. This model could capture the experimental results well with a deviation of less than 5 %. Applying this model, the depth of the liquid water, the temperature of the ambient air and the temperature of the liquid water are considered. When the depth of the liquid water is large enough, the thermal insulation at the bottom of the liquid water is not needed, because the thermal losses dissipating into the liquid water is limited by the convective heat-transfer coefficient of the water at this time. In this work, a depth of 20 mm is large enough. The evaporation efficiency increase linearly with the increase of the temperature of the ambient air or/and the temperature of the liquid water, due to the decreased temperature difference between the bi-layered structure and the ambient air or the liquid water. The temperatures of the liquid water could have a much larger effect on the evaporation efficiency with respect to the ambient air. The influence of the porosity and the thermal conductivity of the second-layer porous material on the evaporation efficiency are also considered. The effective thermal conductivity of the second layer mainly depends on the porosity rather than the thermal conductivity of the second-layer porous material. Thus, we do not need to take into consideration of the thermal conductivity when selecting the second-layer materials.

## Acknowledgment

This work has been supported by the Fundamental Research Funds for the Central Universities (2017XKZD05).